\documentclass[prd,nofootinbib,floats,superscriptaddress,eqsecnum,tightenlines,preprintnumbers,11pt]{revtex4-1}

\usepackage{hyperref}
\usepackage{graphicx}
\usepackage{amsmath,amssymb,amsfonts,amsthm,latexsym,stmaryrd}
\usepackage{marginnote}
\usepackage{xcolor}

\usepackage{physics}
\usepackage{csquotes}
\usepackage{tensor}
\usepackage{mathtools}
\usepackage{amssymb}
\usepackage{siunitx}
\usepackage{caption}
\usepackage{cleveref}
\usepackage{multirow}

\usepackage{rotating}
\usepackage[toc,page]{appendix}
\usepackage{nicefrac}
\usepackage{subcaption}

\def\be{\begin{equation}}
\def\ee{\end{equation}}
\def\beq{\begin{equation}}
\def\eeq{\end{equation}}
\newcommand{\bea}{\begin{eqnarray}}
\newcommand{\eea}{\end{eqnarray}}
\def\bi{\begin{itemize}}
\def\ei{\end{itemize}}
\def\ba{\begin{array}}
\def\ea{\end{array}}
\def\bfig{\begin{figure}}
\def\efig{\end{figure}}

\newcommand\rs{r_{\rm s}}
\newcommand\mass{r_{\rm m}}

\newcommand{\aplus}{a_+}
\newcommand{\aminus}{a_-}

\newcommand\cA{\mathcal{A}}
\newcommand\cB{\mathcal{B}}
\newcommand\cC{\mathcal{C}}
\newcommand\cD{\mathcal{D}}
\newcommand\cF{\mathcal{F}}

\newcommand\fa{d}

\newcommand\cConf{\gamma}
\newcommand\cDis{\varpi}
\newcommand\cCo{\varkappa}

\newcommand{\af}{f_0}
\newcommand{\bb}{f_1}
\newcommand{\gp}{{p_1}}
\newcommand{\rz}{r_-}
\newcommand{\rp}{r_+}

\newcommand{\rg}{r_g}

\begin{document}

\title{
 On the effective metric  of  axial black hole perturbations in DHOST gravity}

\author{David Langlois}
\affiliation{Universit\'e Paris Cit\'e,  CNRS, Astroparticule et Cosmologie, F-75013 Paris, France}

\author{Karim Noui}
\affiliation{Universit\'e Paris-Saclay, CNRS/IN2P3, IJCLab, 91405 Orsay, France}
\affiliation{Universit\'e Paris Cit\'e,  CNRS, Astroparticule et Cosmologie, F-75013 Paris, France}

\author{Hugo Roussille}
\affiliation{Universit\'e Paris Cit\'e,  CNRS, Astroparticule et Cosmologie, F-75013 Paris, France}
\affiliation{Universit\'e Paris-Saclay, CNRS/IN2P3, IJCLab, 91405 Orsay, France}

\date{\today}

\begin{abstract}

We study axial (or odd-parity) perturbations about  static and spherically symmetric hairy black hole (BH) solutions in shift-symmetric DHOST (Degenerate Higher-Order Scalar-Tensor) theories. We first  extend to the family of  DHOST theories the first-order formulation that we  recently developed for Horndeski theories. Remarkably,  we find that  the dynamics of DHOST axial perturbations  is equivalent to  that   of axial perturbations in  general relativity (GR)   evolving in a, distinct,  effective metric. In the particular case
of quadratic DHOST theories, this effective metric is derived  from the background BH metric via a   disformal transformation. We illustrate our general study with three examples of BH solutions. 
In some so-called stealth solutions, the effective metric is Schwarzschild with a shifted horizon. We also give an example of BH solution for which the effective metric is associated with  a naked singularity.
\end{abstract}

\maketitle

\section{Introduction}

With the first direct detections of gravitational waves (GW), the last few years have seen the dawn of  GW astronomy, allowing the study of strongly gravitational systems through the GWs they emit. The archetypal  example of a GW event is a binary black hole merger, which constitutes the vast majority of sources presently observed by GW detectors. The possibility to observe these systems directly gives hope that one might be able to reach a precision sufficient  enough to compare observations with the predictions of General Relativity (GR) and put constraints on deviations from GR in strong gravity regimes. This  perspective puts modified theories of gravity in the limelight, since  they are useful  to test GR, by comparison of their respective predictions. 
  Therefore, much work has been done in the past few years in the direction of new theories of gravity and finding new black hole (BH) solutions in these theories. Once a solution is obtained, it is important to study its perturbations in order to compute the behaviour of GWs and to make sure no theoretical issues that could rule out the solution are present, before making any prediction for experimental results.

In this paper, we focus our attention on  axial (or odd-parity) perturbations of nonrotating black holes in the context of  Degenerate Higher-Order Scalar-Tensor (DHOST) theories \cite{Langlois:2015cwa,Langlois:2015skt,BenAchour:2016cay,Crisostomi:2016czh,BenAchour:2016fzp}, the most general family of scalar-tensor theories propagating a {\it single  scalar} degree of freedom. 
In DHOST theories, axial perturbations  involve, as in GR,  a single degree of freedom,  in contrast with polar  (or even-parity) perturbations which contain two degrees of freedom, one from the metric and the other from the scalar field. For this reason, axial perturbations are easier to study and they have already been investigated in several works in the context of DHOST theories, or subfamilies: from generic BH backgrounds  in   
\cite{Kobayashi:2012kh,Cisterna:2015uya,Takahashi:2016dnv, Takahashi:2019oxz,Chatzifotis:2021pak}
to \enquote{stealth} solutions (solutions with a nontrivial scalar hair but whose metric coincides  with the GR Schwarzschild metric)  in \cite{Tomikawa:2021pca}. 
{Other works also include polar perturbations~\cite{deRham:2019gha,Khoury:2020aya,Takahashi:2021bml,Kase:2021mix,Minamitsuji:2022mlv,Minamitsuji:2022vbi}.
}
Many of these studies rely  on the computation of the quadratic Lagrangian that describes the dynamics of the perturbations. Recently, we proposed another approach based on the extraction of a first-order system from the perturbed Einsteins' equations  in \cite{Langlois:2021xzq,Langlois:2021aji}. Note that a few other works have relied on  an EFT approach  \cite{Franciolini:2018uyq,Hui:2021cpm,Mukohyama:2022enj}.

The present work focuses on this first-order formulation of axial perturbations  and uses it to compute the effective metric in which axial perturbations propagate. We start by recalling the structure of  DHOST theories and describe a few known black hole solutions for these theories, namely stealth solutions \cite{Babichev:2013cya}, the BCL solution introduced in \cite{Babichev:2017guv} and the 4-dimensional Einstein-Gauss-Bonnet (4dEGB) solution proposed in \cite{Lu:2020iav}. We then extend the first-order formalism developped  for quadratic Horndeski theories  in \cite{Langlois:2021aji}  to  the full family of DHOST theories (defined up to cubic order in second derivatives of the scalar field). For axial perturbations, which involve a single degree of freedom, one can reexpress the equations of motion in the form of  a Schr\"odinger-like  equation for an appropriate master variable.

 We then show that axial perturbations verify the equation of propagation of a massless 
 spin-2 field in GR, on a background described by some effective metric that depends on  the functions in the DHOST Lagrangian  and the BH geometry. 
 In the case of quadratic DHOST theories, we find that  the effective metric can be obtained directly  from the background solution via a disformal metric.
 We illustrate our analysis by applying it to the particular BH solutions mentioned above. 
 
 In particular,  we discuss stealth solutions  whose effective metric is  Schwarzschild but  with a shifted horizon in general. This case was also  studied recently in \cite{Tomikawa:2021pca, Takahashi:2021bml}.  Effectively, axial perturbations and non-gravitational fields (photons, matter) propagate in different metrics since non-gravitational fields are minimally coupled to the metric. It is thus important to check that the causal structures are compatible.
 Despite having different horizons for the axial perturbations and the background, we find that the lightcones associated to both are compatible\footnote{The polar sector, which we do not consider in the present work, is however pathological, as stressed in  \cite{deRham:2019gha}.  A possible way out involves a slight detuning of the degeneracy conditions \cite{Motohashi:2019ymr},  which for example manifests itself in U-DHOST theories \cite{DeFelice:2018ewo}, as discussed in \cite{DeFelice:2022xvq}.}. We also consider the effective metric of the 4dEGB solution's axial perturbations and find that it is not a BH metric but instead a naked singularity, which 
 is consistent with  the pathological asymptotic behaviours found for this solution in \cite{Langlois:2022eta}. 

The outline of the paper is the following. In Sec.~\ref{sec:axial-cubic-DHOST}, we give a brief presentation of  DHOST theories and of the few BH solutions that will illustrate our analysis. We then turn, in Sec.~\ref{sec:axial-dynamics}, to the equations of motion for the axial perturbations, using the first-order framework  and then the Schr\"odinger-like approach.
 In Sec.~\ref{sec:effective-metric}, we compute the effective metric for axial perturbations from different perspectives. We then apply these general results to  our few examples. This article is completed with a few appendices, in which we give additional details and  provide some connections with  previous works.

\section{DHOST theories and examples of BH solutions}
\label{sec:axial-cubic-DHOST}

In this section, we briefly present the family of DHOST theories, introduced in  \cite{Langlois:2015cwa,Langlois:2015skt} and extended to cubic order (in second derivatives of the scalar field) in \cite{BenAchour:2016fzp}  (see also \cite{Langlois:2018dxi} for a review). We then describe of few exacts BH solutions in the context of DHOST theories.

\subsection{DHOST theories}

We consider the family of DHOST theories, up to cubic order, whose   action written in terms of the metric $g_{\mu\nu}$  and  the scalar field $\phi$  takes the form
\begin{eqnarray}
	S[g_{\mu\nu},\phi] = \int \dd[4]{x} \sqrt{-g} \Big(P(X,\phi) + Q(X,\phi) \square\phi + L^{(2)} + L^{(3)} \Big) \, ,
	\label{eq:generic-quintinc-Horndeski}
\end{eqnarray}	
where $X\equiv  \nabla^\mu\phi\, \nabla_\mu\phi$ and the $L^{(2)}$ and $L^{(3)}$ denote, respectively, the quadratic and cubic contributions, including the associated curvature-dependent terms, 
\begin{equation}
	L^{(2)} = F_2(X,\phi) R + \sum_{i=1}^5 {A}_i(X,\phi) L_i^{(2)} \,, \quad 
	L^{(3)} = F_3(X,\phi) G_{\mu\nu} \phi^{\mu\nu} + \sum_{i=1}^{10} B_i(X,\phi) L_i^{(3)} \,.
\end{equation}
In the above expressions, 
$G_{\mu\nu}$  is the Einstein tensor, $R$ the Ricci scalar, and we use the shorthand notation $\phi_\mu \equiv \nabla_\mu \phi$ and $\phi_{\mu\nu}\equiv \nabla_\mu \nabla_\nu \phi$  for the first and second (covariant) derivatives of the scalar field.
The five elementary quadratic Lagrangian  $L_i^{(2)}$, introduced in  \cite{Langlois:2015cwa}, read
\begin{align}
&L^{(2)}_1 = \phi_{\mu \nu} \phi^{\mu \nu} \,, \qquad
L^{(2)}_2 =(\Box \phi)^2 \,, \qquad
L_3^{(2)} = (\Box \phi) \phi^{\mu} \phi_{\mu \nu} \phi^{\nu} \,,  \nonumber \\
&L^{(2)}_4 =\phi^{\mu} \phi_{\mu \rho} \phi^{\rho \nu} \phi_{\nu} \,, \qquad
L^{(2)}_5= (\phi^{\mu} \phi_{\mu \nu} \phi^{\nu})^2\, ,
\end{align}
while the ten  elementary cubic Lagrangian densities $L_i^{(3)}$  are given by \cite{BenAchour:2016fzp}
\begin{align}
&L^{(3)}_1=  (\Box \phi)^3  \,, \quad
L^{(3)}_2 =  (\Box \phi)\, \phi_{\mu \nu} \phi^{\mu \nu} \,, \quad
L^{(3)}_3= \phi_{\mu \nu}\phi^{\nu \rho} \phi^{\mu}_{\rho} \,,  \nonumber \\
&L^{(3)}_4= \left(\Box \phi\right)^2 \phi_{\mu} \phi^{\mu \nu} \phi_{\nu} \,, \quad
L^{(3)}_5 =  \Box \phi\, \phi_{\mu}  \phi^{\mu \nu} \phi_{\nu \rho} \phi^{\rho} \,, \quad
L^{(3)}_6 = \phi_{\mu \nu} \phi^{\mu \nu} \phi_{\rho} \phi^{\rho \sigma} \phi_{\sigma} \,, \nonumber  \\
&L^{(3)}_7 = \phi_{\mu} \phi^{\mu \nu} \phi_{\nu \rho} \phi^{\rho \sigma} \phi_{\sigma} \,, \quad
L^{(3)}_8 = \phi_{\mu}  \phi^{\mu \nu} \phi_{\nu \rho} \phi^{\rho}\, \phi_{\sigma} \phi^{\sigma \lambda} \phi_{\lambda} \,,  \nonumber \\
&L^{(3)}_9 = \Box \phi \left(\phi_{\mu} \phi^{\mu \nu} \phi_{\nu}\right)^2  \,, \quad
L^{(3)}_{10} = \left(\phi_{\mu} \phi^{\mu \nu} \phi_{\nu}\right)^3 \,.
\end{align}

From now on,  we restrict our study to shift-symmetric theories, which entails that 
all the  functions  in the action \eqref{eq:generic-quintinc-Horndeski}  depend on the kinetic density  $X$ only.
Although we do not write explicitly the action that governs the non-gravitational fields, we stress that these fields are assumed to be minimally coupled to the metric $g_{\mu\nu}$. This will be an important point later in this paper when we compare the metrics effectively ``seen" by the gravitational perturbations and non-gravitational fields, respectively.

While  the functions $P$ and $Q$ can be chosen arbitrarily, the other ones, $F_i$, $A_i$ and $B_i$, must satisfy  degeneracy conditions in order to guarantee the presence of a single scalar degree of freedom. These degeneracy conditions were explicitly computed in
\cite{BenAchour:2016fzp}, generalising the degeneracy conditions for quadratic DHOST theories established in \cite{Langlois:2015cwa}.

In the case of  quadratic DHOST theories, it was shown that    physically viable theories belong to the class Ia (or ${}^2$N-I as per the classification of \cite{Crisostomi:2016czh}) where the functions $F_2$, $A_1$ and $A_3$ are free while the two others are related to these first ones by the degeneracy conditions.
In the case of cubic DHOST theories, the most relevant theories belong to the class ${}^3$N-I where only $F_3$, $B_2$ and $B_6$ are free
while the others depend on these three functions \cite{BenAchour:2016fzp}. In particular, we have the conditions
\begin{eqnarray}
3B_1+B_2 \, = \, 3 B_3 + 2 B_2 \, = \, 0 \,.
\label{eq:degen-DHOST-cubic}
\end{eqnarray}
If one allows for both quadratic and cubic terms,  for instance theories in ${}^2$N-I and theories in ${}^3$N-I, one has to consider new degeneracy
conditions, which link quadratic and cubic functions as follows:
\begin{eqnarray}
X  F_2  B_6 & = & A_1 X F_{3X} - 2 B_2 F_2 + 2B_2 X F_{2X} - 2 F_2 F_{3X} \, , 
\label{degeneracy1}
\\
 X^2 F_2 B_2 A_3 & = & 2 B_2 (3 X A_1  F_2 - 4  X^2 A_1 F_{2X} + 2 XF_2 F_{2X} - 2 F_2^2) 
- 4 F_{3X}(F_2 - X A_1)^2 \,,
\label{degeneracy2}
\end{eqnarray}
where  $F_X$ denotes  the derivative of any function $F$ with respect to $X$.
The first relation fixes $B_6$ (when $F_2 \neq 0$) and the second one fixes $A_3$  (when $F_2 B_2 \neq 0$). 
Hence, as the classes ${}^2$N-I and  ${}^3$N-I are independently parametrised by 3 free functions each, the ``merged" class that  combines
 them depends on four free functions due to the  two degeneracy conditions \eqref{degeneracy1}-\eqref{degeneracy2}.
 
  It was shown in \cite{BenAchour:2016fzp}  that any theory in the merged class is related to Horndeski theories by a disformal transformation\footnote{Note however that in the Horndeski frame, i.e. with the metric for which gravitation is described by a Horndeski action, the standard matter fields are no longer minimally coupled to this metric. The interest of the DHOST formulation is precisely that the matter fields can be assumed to be minimally coupled to the metric without loss of generality.}. As Horndeski theories
depend on two free functions and disformal transformations are parametrized by two others functions, we recover that the merged class
involves four free functions. 
As discussed in 
\cite{Langlois:2017mxy}, all  DHOST theories that are related to Horndeski theories via disformal transformations belong to a special  category, which was named  ${\cal C}_I$, in contrast with the other classes of DHOST theories which form the category ${\cal C}_{II}$ and are not physically interesting.

\subsection{A few BH solutions}
\label{subsection:BH_solutions}
In the rest of this paper, we will be interested in  static and spherically symmetric  solutions of these theories,  characterised by  a metric  
\begin{equation}
\label{metric}
	\dd{s}^2 = - \cA(r) \dd{t}^2 + \frac{1}{\cB(r)} \dd{r}^2 + \cC(r)  \dd{\Omega}^2 \,, \qquad
	 \dd{\Omega}^2 = \dd\theta^2 + \sin^2\theta \, \dd\varphi^2 \, ,
\end{equation}
and a scalar field of the form
\begin{eqnarray}
\label{generalscalar}
\phi(t,r) = q t + \psi(r) \, ,
\end{eqnarray}
as initially proposed in  \cite{Babichev:2013cya}, 
where $q$ is a constant and $\psi$ is a function of $r$. Note that the time dependence disappears in the gradient of $\phi$.

Various exact BH solutions of this type have been obtained~\cite{Babichev:2017guv,BenAchour:2018dap,Motohashi:2019sen,Minamitsuji:2019shy,BenAchour:2020wiw,Minamitsuji:2019tet,Takahashi:2020hso} (see also the reviews \cite{Babichev:2016rlq,Kobayashi:2019hrl} on Horndeski theories and references therein). In this work, we are going to focus our attention on  three particular solutions, which we briefly describe below. 

\subsubsection*{Stealth solutions}
Stealth solutions are solutions in which the metric coincides with a vacuum solution of GR, possibly with a cosmological constant. They have been actively studied, since by construction they allow one to recover many predictions of GR. If, in addition one assumes a constant kinetic density $X$, such solutions exist when the theory  satisfies a set of conditions that have been  listed in \cite{Minamitsuji:2019shy} for DHOST theories. Specific solutions were found in \cite{Babichev:2013cya, Babichev:2016kdt, Babichev:2017lmw, BenAchour:2018dap, Charmousis:2019vnf}. In particular, ${\cal C}_{I}$ theories  we have described above can be shown to admit a stealth solution 
where $X=-q^2$ and the metric is the Schwarzschild solution, i.e.
\begin{eqnarray}
\label{metricstealth}
\cA(r)=\cB(r) = 1 - \frac{\rs}{r} \, , \qquad \cC(r)=r^2 \, ,
\end{eqnarray}
with $\rs$ the black hole horizon radius, when
\begin{eqnarray}
P=P_X=Q_X=B_2=0 \quad {\rm for}\ X=-q^2\,.
\end{eqnarray}
All the other conditions given in Eq. (18) of \cite{Minamitsuji:2019shy} are trivially satisfied for  ${\cal C}_I$ theories. As for the scalar field \eqref{generalscalar}, it is obtained by integrating the equation
\begin{eqnarray}
\label{scalarstealth}
 \psi'(r) = \frac{q \sqrt{\rs r}}{r-\rs} \, . 
\end{eqnarray}

Although much studied, stealth solutions also appear to suffer from pathologies,  as pointed out in \cite{deRham:2019gha} for instance. 
Non stealth solutions are more complicated to find and very few have been constructed so far. In this work, we consider two examples of non-stealth solutions. 

\subsubsection*{BCL solution}
The first one, dubbed  BCL (for Babichev-Chamousis-Lehebel)  was obtained in \cite{Babichev:2017guv} for a subset of quadratic Horndeski theories  with
\begin{eqnarray}
F_2(X)= f_0 + f_1 \sqrt{X} \, , \qquad P(X)= - p_1 X \, , \qquad Q(X)=0 \, ,
\end{eqnarray}
where  the coefficients $f_0$, $f_1$ and $p_1$ are constant (with $f_0>0$ and $p_1>0$). Notice that 
such a theory exists only if $X>0$, which we are going to assume. The BCL solution has been
obtained for any value of $q$ but, for simplicity, we will only consider the case $q=0$ here. In this case, the scalar field is time-independent and the  black hole solution found in \cite{Babichev:2017guv}  simplifies: it is described by a  metric of the form \eqref{metric} with 
 \beq
 \label{BCLrprm}
\cA(r) = \cB(r) =  \left(1-\frac{r_+}{r} \right)\left(1+\frac{r_-}{r} \right) \qq{and} \cC(r) = r^2 \,,
\eeq
where the two radii $\rz$ and $\rp$ satisfy
\bea
\label{rprm}
r_+ r_- = \frac{\bb^2}{2\af \gp} \, , \qquad r_+ - r_- =\mass  \, \equiv  \, 2m \, , \qquad 0 < r_- <  r_+ \,,
\eea
$m$ being the BH mass.
One can see that the metric described in \eqref{BCLrprm} is very similar to that of a Reissner-Nordstr\"om  BH. However, since $r_->0$, the BCL black hole exhibits only a single event horizon, of radius $r_+$, in contrast with the Reissner-Nordstr\"om geometry. 
The scalar field can be found explicitly and is given by
\beq
\label{BCLscalar}
\phi(r)=\psi(r)=\pm \frac{\bb}{\gp\sqrt{\rp\rz}}\arctan\left[\frac{\mass r + 2\rp\rz}{2\sqrt{ \rp\rz}\sqrt{(r-\rp)(r+\rz)}}\right]\, + {\text{cst}} \, .
\eeq
The global sign of $\phi(r)$ and the constant are physically irrelevant \cite{Babichev:2017guv}. 

\subsubsection*{$D\rightarrow4$ Gauss-Bonnet solution} 

The second non-stealth solution we will discuss  was found in  \cite{Lu:2020iav} where the authors considered  a specific limit $D \rightarrow 4$ of the $D$-dimensional Einstein-Gauss-Bonnet action. In that way, they obtained a solution of a particular  Horndeski theory  defined by
\begin{equation}
\label{functions_4dGB}
	P(X) = 2\alpha X^2 \,,\quad Q(X) = -4\alpha X \,,\quad F_2(X) = 1 - 2\alpha X \qq{and} F_3(X) = -4\alpha \ln X \,,
\end{equation}
where $X$ is supposed to be positive and $\alpha>0$ is a  constant of the theory with the dimension of a length squared. In the limit $\alpha \rightarrow 0$, one recovers the Einstein-Hilbert action. The solution found in \cite{Lu:2020iav,Hennigar:2020lsl} is static and spherically symmetric with a metric given by 
\begin{eqnarray}
	\cA(r) =\cB(r) = 1-  \frac{2\rs}{r(1+\sqrt{1+{4\alpha\rs}/{r^3}})}\,, 
\qquad  \cC(r)=r^2
\,. \label{eq:metric-function-r}
\end{eqnarray}
The solution reduces to the Schwarzschild metric in the limit   $\alpha \rightarrow 0$,  as expected, with the integration constant $\rs$  corresponding to the BH horizon radius. When $\alpha$ is not too large, more precisely $\alpha \leq \rs^2/4$, the solution still describes a BH geometry. It admits two (inner and outer)  horizons located at the positions
\begin{equation}
	r_\pm = \frac{1}{2} \left( \rs \pm {\sqrt{\rs^2-4\alpha}} \right) \,,
\end{equation}
obtained  by solving $\cA(r)=0$.
If $\alpha>\rs^2/2$, the solution is a naked singularity, which is not interesting for us.

As for  the scalar field, it is time-independent, i.e. $q=0$, and its radial profile is given by\footnote{We consider here only the most interesting branch from a physical point of view. The other branch is also discussed in \cite{Langlois:2022eta}.}
\begin{equation}
	\phi'(r) = \frac{\sqrt{\cA(r)}-1}{r \sqrt{\cA(r)}} \, .
	\label{eq:sol-phi}
\end{equation}
The scalar field behaves as  $\phi(r) \simeq \rs/(2 r)$ when $r \gg \rs$  and $\phi(r) \propto \sqrt{r-r_s}$ when $|r-\rs| \ll \rs$.

\section{Dynamics of axial perturbations}
\label{sec:axial-dynamics}
In this section, we consider the linear axial perturbations about the generic background solution \eqref{metric} and \eqref{generalscalar}, and we briefly recall how to compute their equations of motion following the methods and notations of \cite{Langlois:2021aji} (see also the general reviews \cite{Kokkotas:1999bd, Nollert:1999ji, Berti:2009kk, Konoplya:2011qq}  on quasi-normal modes and  \cite{Kobayashi:2012kh,Kobayashi:2014wsa} for the first study of BH perturbations  in Horndeski theories). 

We  define the metric perturbations $h_{\mu\nu}$ by
\begin{equation}
	g_{\mu\nu} = \bar{g}_{\mu\nu} + h_{\mu\nu} 
		\,,
\end{equation}
where a bar denotes the background solution. We restrict ourselves to the study of axial perturbations in the Regge-Wheeler gauge (see the original paper \cite{Regge:1957td} and the more recent article \cite{Kobayashi:2012kh} for the notations we are using here) where the only non-vanishing components of the perturbations are
\begin{align}
	&h_{t\theta} = \frac{1}{\sin\theta}  \sum_{\ell, m} h_0^{\ell m}(t,r) \partial_{\varphi} {Y_{\ell m}}(\theta,\varphi), \qquad
	h_{t\varphi} = - \sin\theta  \sum_{\ell, m} h_0^{\ell m}(t,r) \partial_{\theta} {Y_{\ell m}}(\theta,\varphi), \nonumber \\
	&h_{r\theta} =  \frac{1}{\sin\theta}  \sum_{\ell, m} h_1^{\ell m}(t,r)\partial_{\varphi}{Y_{\ell m}}(\theta,\varphi), \qquad
	h_{r\varphi} = - \sin\theta \sum_{\ell, m} h_1^{\ell m}(t,r)  \partial_\theta {Y_{\ell m}}(\theta,\varphi), \label{eq:odd-perttext}
\end{align}
using an expansion in spherical harmonics ${Y_{\ell m}}(\theta,\varphi)$  because of the spherical symmetry of the background. In the following, since perturbations with different values of $\ell$ and $m$ do not couple at the linear level, we drop the indices $\ell$ and $m$ for clarity. It will also be convenient to  use 
\begin{equation}
 \lambda = \frac{\ell (\ell +1)}{2} - 1 \, ,
\label{lambda}
\end{equation}
which will contain the whole dependence on $\ell$.

The equations of motion are obtained from the action \eqref{eq:generic-quintinc-Horndeski}  expanded at  quadratic order  in the perturbations $h_{\mu\nu}$.
Note that the scalar field perturbation $\delta\phi$ vanishes identically when one considers axial perturbations.

\subsection{First order differential system}
A long but straightforward calculation, whose details are given in App.~\ref{Appendix:Axial}, shows that the dynamics of 
the perturbations  $h_0$ and $h_1$ can be described in terms of a   differential  system of the form
\begin{eqnarray}
\pdv{Y}{r} + \Psi \pdv{Y}{t} = M_0 \, Y + M_2  \, \pdv[2]{Y}{t} \, ,
\label{eq:system-axial-canonical}
\end{eqnarray}
where we have introduced the vector $Y$ defined by
\begin{equation}
Y = \left( \begin{array}{c} Y_1 \\ Y_2 \end{array}\right) \qquad \text{with} \qquad Y_1=h_0\, , \quad
	\omega Y_2 = h_1 + \Psi h_0  \,,
\end{equation}
while $M_0$ and $M_2$ are two matrices given by
\begin{eqnarray}
M_0 = \begin{pmatrix}
		 {\cC'}/{\cC} & {2 i \lambda \Phi}/{\cC } \\
		- i \Gamma & \Delta 
	 \end{pmatrix} \, , \qquad M_2 =  
	 \begin{pmatrix}
	 0 & i \\
	 0 & 0 
	 \end{pmatrix}   \, .
\end{eqnarray}
The different functions $\Phi$, $\Psi$, $\Gamma$ and $\Delta$ entering  the system depend on the background solution and also on the DHOST theory functions \eqref{eq:generic-quintinc-Horndeski} evaluated on the background. They can be deduced immediately from the  following relations:
\begin{align}
\mathcal{F} &= \begin{multlined}[t] \cA F_2 -  (q^2 + \cA X) A_1 - \frac12 \cA \cB \psi'  X' F_{3X} - \frac12 \cB \psi' (\cA X)'  B_2 -\frac{\cA}{2\cB} (\cB \psi')^3 X' B_6\,,\end{multlined}\nonumber\\
	\mathcal{F} \Psi &= q \left[\psi' A_1 + \frac 12  \left(\cB \psi^{\prime 2} \right)' F_{3X} + \frac{1}{2} \frac{(\cA X)'}{\cA}  B_2 + \frac 14 \left(\cB^2 \psi^{\prime 4} \right)'B_6\right] \,,\nonumber\\
	\frac{\mathcal{F}}{\Phi} &= F_2 - X A_1 - \frac12 \cB \psi^{\prime} X' F_{3X}  - \frac12 \cB \psi^{\prime}  \frac{(\cC X)'}{\cC } B_2 - \frac12 \cB \psi^{\prime} XX' B_6\,,\nonumber\\
	\Gamma &= \Psi^2 + \frac{1}{2\cA\cB\mathcal{F}} \qty( 2{q^2 A_1} + 2\cA {F_2} +  \cA \cB \psi^{\prime} X' F_{3X}  + {q^2} \frac{(\cA X)' }{\cA  \psi^{\prime}} B_2 + {q^2} \cB \psi^{\prime} X' B_6) \,,\nonumber\\
	\Delta &= - \frac{\mathcal{F} '}{\mathcal{F} } - \frac{\cB'}{2\cB} + \frac{\cA'}{2\cA}\,,
	\label{PhiGammaDeltaPhi}
\end{align}
where a prime denotes a derivative  with respect to $r$ and we have imposed the degeneracy condition $3 B_3 + 2 B_2 = 0$ from \eqref{eq:degen-DHOST-cubic} to simplify the expressions. One can note that only the functions $F_2$, $F_3$, $A_1$, $B_2$ and $B_6$ appear in the perturbations: this could be expected from the ADM decomposition of the DHOST action \eqref{eq:generic-quintinc-Horndeski} given in \cite{Langlois:2017mxy}, since only the terms containing contractions of the extrinsic curvature tensor of the form $K^{ij}K_{ij}$ or $K^{ij} K_{jk} \tensor{K}{_k^i}$ contain couplings with the axial modes (this can be understood by looking at the quadratic actions for tensor modes  given in \cite{Gleyzes:2014rba} and \cite{Langlois:2017mxy}). In the case of GR (where $F_2=1$ and $F_3=A_1=B_2=B_6=0$),  \eqref{PhiGammaDeltaPhi} boils down to
\begin{equation}
	\mathcal{F} = \cA \,,\quad \Psi = 0 \,,\quad \Phi = \cA \,,\quad \Gamma = \frac{1}{\cA \cB} \qq{and} \Delta = -\frac{\cA'}{2\cA} - \frac{\cB'}{2\cB} \,.
	\label{eq:PhiGammaDeltaPhi-GR}
\end{equation}

\medskip

In the general case, we can slightly simplify the differential system  \eqref{eq:system-axial-canonical}  and absorb the first time derivative of $Y$ with the following change of  time coordinate
\begin{eqnarray}
\label{defoftstar}
 t_* = t - \int \mathrm{d}r \,  \Psi(r) \, ,
\end{eqnarray}
which transforms  \eqref{eq:system-axial-canonical} into
\begin{eqnarray}
\pdv{Y}{r} = M_0 \, Y + M_2  \, \pdv[2]{Y}{t_*} \, .
\end{eqnarray}
Hence, one can  eliminate the function $\Psi$ from the differential system \eqref{eq:system-axial-canonical} by a simple redefinition of the time coordinate.

Then, we expand in Fourier modes and, using the convention 
\begin{eqnarray}
\label{Fouriertstar}
f(t_*,r)=e^{-i\omega t_*} f(r)
\end{eqnarray} 
for any function $f$,  we obtain the first-order differential system in the radial variable:
\begin{equation}
	\dv{Y}{r}= M Y \, , \quad \text{with} \quad M=
	\begin{pmatrix}
		 {\cC'}/{\cC} & -i\omega^2 + {2 i \lambda \Phi}/{\cC } \\
		- i \Gamma & \Delta 
	 \end{pmatrix} \,.
	 \label{firstordertstar}
\end{equation}
This system generalizes the first-order system obtained for axial perturbations in quadratic Horndeski theories, given in \cite{Langlois:2021aji}.

To summarize, axial perturbations about a general static and spherically symmetric background \eqref{metric} in DHOST theories are fully described in terms of the previous first-order system where the time coordinate $t_*$ is related to the original time coordinate $t$ that appears in the background metric by the transformation \eqref{defoftstar}. 
\medskip

If we assume the action to be that of GR,   the previous  dynamical system \eqref{firstordertstar}  simplifies and, using \eqref{eq:PhiGammaDeltaPhi-GR},   the matrix $M$ reduces to 
\begin{equation}
	 M_\text{GR}=
	\begin{pmatrix}
		 {\cC'}/{\cC} & -i\omega^2 + {2 i \lambda}\cA/{\cC } \\
		- i /(\cA\cB) & -(\cA'/\cA+\cB'/\cB)/2
	 \end{pmatrix} \,,
	 \label{MatrixMspin2}
	\end{equation}
for an arbitrary background metric. Note that the background metric  is not necessarily a {\it vacuum} solution of GR.  It can also be any  solution of Einstein's equations with an energy-momentum tensor  whose perturbations vanish in the axial sector.

\subsection{Schr\"odinger-like equation}
In this subsection, we show how to recover a Schr\"odinger-like equation for the axial perturbations starting from the first order system  \eqref{firstordertstar}. Our results  are consistent with those obtained recently from the quadratic action directly in \cite{Tomikawa:2021pca,Takahashi:2021bml} for instance.

The two first-order equations in \eqref{firstordertstar}, namely
\begin{eqnarray}
\dv{Y_1}{r}=  \frac{\cC'}{\cC} Y_1  +i \qty(\frac{2 \lambda \Phi}{\cC }- \omega^2 ) Y_2 \, , \qquad \dv{Y_2}{r}= - i \Gamma Y_1 + \Delta Y_2 \,,
\end{eqnarray}
can be combined into a single second-order equation, which reads
\begin{eqnarray}
\dv[2]{Y_2}{r} - \left(\Delta + \frac{\Gamma'}{\Gamma} + \frac{\cC'}{\cC} \right) \dv{Y_2}{r} + \left[\Gamma \left( \omega^2 - 2 \lambda \frac{\Phi}{\cC}\right) + \Delta \left( \frac{\Gamma'}{\Gamma} + \frac{\cC'}{\cC}\right) \right]Y_2=0 \, .
\end{eqnarray}
To obtain a Schr\"odinger-like equation, one needs to get rid of the first derivative term.  This can be done either by introducing a new radial coordinate  $r_*$, via the reparametrisation
\begin{eqnarray}
\label{defofrstar}
\dv{r}{r_*} = n(r) \, , 
\end{eqnarray}
or by a rescaling of the function $Y_2$,
\begin{eqnarray}
Y_2(r)=N(r) \, {\cal Y}(r) \,.
\end{eqnarray}
Combining both transformations for generality,  and imposing the condition
\begin{eqnarray}
\label{RelationN}
2 \frac{N'}{N} - \frac{n'}{n} = \Delta + \frac{\Gamma'}{\Gamma} + \frac{\cC'}{\cC} \,,
\end{eqnarray}
or, equivalently, that $N^2/n$ is proportional to $\cA^{1/2} \cC\, \Gamma/(\cB^{1/2} \mathcal{F})$, one gets the Schr\"odinger-like equation 
\begin{eqnarray}
\label{Schrodingerequation}
\dv[2]{{\cal Y}}{r_*}  + \left( \frac{\omega^2}{c_*^2(r)}  - V(r) \right) {\cal Y} = 0 \, .
\end{eqnarray}
The propagation speed  $c_*(r)$, associated with the coordinate system $(t_*,r_*)$,  is defined by the relation
\begin{eqnarray}
\label{speedcstar1}
 c_*^2 = \frac{1}{n^2 \Gamma} \, ,
\end{eqnarray}
while the potential is given by
\begin{eqnarray}
\label{Veff}
{V}=  n^2\left(2 \lambda \frac{\Gamma \Phi}{\cC} + V_0 \right)\, , 
\end{eqnarray}
with 
\begin{eqnarray}
\label{V_0}
{V}_0 &= & \frac{1}{4} \Big[ \Delta^2 + 2  \Delta' - 2 \Delta \left( \frac{\Gamma'}{\Gamma} + \frac{\cC'}{\cC}\right)
+ 2 \frac{ \Gamma' \cC'}{\Gamma \cC}  \nonumber \\
&& \qquad +3 \left( \frac{\Gamma'}{\Gamma}\right)^2 + \left( \frac{n'}{n}\right)^2 + 3  \left(\frac{\cC'}{\cC}\right)^2 - 2 \left(  \frac{\Gamma''}{\Gamma} +  \frac{n''}{n} +  \frac{\cC''}{\cC} \right) \Big] \, .
\end{eqnarray}
It is sometimes interesting to choose a coordinate system where  the propagation speed is normalized to  $c_*=1$. In that case, the radial reparametrisation function $n$ is fixed
by the relation  $n=1/\sqrt{ \Gamma}$, according to \eqref{speedcstar1},  and the expression of $V$ simplifies slightly.  

\medskip

In the special case of GR,  choosing $n=1/\sqrt{ \Gamma}=\sqrt{\cA\cB}$, the expression of the potential  \eqref{Veff} reduces to
\begin{eqnarray}
\label{PotABC}
V =  2\lambda \frac{\cA}{\cC}+ \frac{1}{2}  \frac{\cD^2 {\cC'}^2}{\cC} - \frac{1}{2} \cD \left(\cC' \cD\right)' \qq{with} \cD = \sqrt{{\cA\cB}/{\cC}} \, ,
\end{eqnarray}
which is --- as expected --- the potential for a massless spin 2 field propagating in the metric \eqref{metric}  
(see \cite{Chandrasekhar:1985kt} and also the recent paper \cite{Arbey:2021jif} which presents a pedagogical review on the dynamics of fields with different spins in black holes using the Newman-Penrose formalism). Note that \eqref{PotABC} reduces to the usual Regge-Wheeler potential when one substitutes the Schwarzschild expressions \eqref{metricstealth}  for $\cA$, $\cB$ and $\cC$.

\section{Effective metric for axial perturbations}
\label{sec:effective-metric}
 In this section,  we show that the dynamics of axial perturbations in the static and spherically symmetric metric  \eqref{metric} 
 is equivalent to the GR dynamics of the axial component of a massless spin 2 field \eqref{MatrixMspin2} propagating in an effective metric $\tilde{g}_{\mu\nu}$ that we compute explicitly.

 \subsection{Effective Klein-Gordon equation}
 \label{section:conformal}
 A simple way to determine the effective metric,  but only up to a global factor, is to interpret the Schr\"odinger-like equation as an effective Klein-Gordon equation of the form
 \beq
 \label{KG}
 \tilde g^{\mu\nu}\, \tilde\nabla_\mu\tilde\nabla_\nu \chi -m^2_{\rm eff}\chi=0\, ,
 \eeq
 where $\tilde\nabla$ denotes the covariant derivative associated with the effective metric $\tilde g_{\mu\nu}$.

Starting from  \eqref{Schrodingerequation}, one simply needs to restore the explicit dependence on  time and angular coordinates, by introducing
\begin{eqnarray}
\chi(t_*,r,\theta,\varphi) \; = \; e^{-i\omega t_*}  {\cal Y}(r) \, Y_{\ell,m}(\theta,\varphi)  \, ,
\label{fullchi}
\end{eqnarray}
which leads, after division of the  Schr\"odinger-like equation  by $\Gamma\Phi$,  to the expression
\begin{eqnarray}
\label{eq_chi}
-   \frac1\Phi \pdv[2]{\chi}{t_*} + \frac{1}{n^2\Gamma\Phi}\pdv[2]{\chi}{r_*}  +  \frac{1}{\cC}    \Delta\!^{(2)} \chi -  \left( \frac{V_0}{\Gamma \Phi} -  \frac{2}{\cC}\right)\chi = 0 \, ,
\end{eqnarray}
where $ \Delta\!^{(2)}$ is the spherical Laplace operator.
By comparison with \eqref{KG}, one can immediately identify the coefficients of the inverse metric $\tilde g^{\mu\nu}$, up to a global factor. The effective metric is thus of the form
\begin{eqnarray}
\label{conformalmetric-tstar}
\dd{\tilde s}^2 = \tilde{g}_{\mu\nu} \dd{x^\mu} \dd{x^\nu} = \Xi \left[- \Phi  \dd{t_*}{}\!\!\!^2 +  \Gamma \Phi \, n^2 \dd{r_*}{}\!\!\!^2 + {\cC} \dd{\Omega}^2 \right] \, .
\end{eqnarray}
This result can also be derived directly from the first-order system as shown in App. \ref{dispersionfirstorder}.

Going back to the original coordinate system, the effective metric thus reads
\begin{eqnarray}
\label{conformalmetric}
\dd{\tilde s}^2  =\Xi \left[- \Phi(\dd{t} - \Psi \dd{r})^2 + {\Gamma \Phi} \dd{r}^2 + \cC \dd{\Omega}^2 \right]\, .
\end{eqnarray}
An interesting question is, depending on the BH background solution under consideration, whether this effective metric describes or not a BH geometry too, with the possibility that the effective horizon might differ from the background horizon. We will see that our examples provide different answers, without exhausting the question.

  \subsection{Equivalence with GR axial perturbations}
 In this subsection, we go one step further and show that the dynamics of the DHOST axial perturbations in the background is equivalent to that  of GR axial perturbations in an effective background \eqref{conformalmetric} with a {\it specific}  conformal factor $\Xi$.
 
 Our starting point is the DHOST system  \eqref{firstordertstar}, which can be rewritten, via a change of vector $\tilde Y = \cConf(r) \, Y$ where $\cConf$ is some function of $r$,  in the form 
 \begin{equation}
	\dv{\tilde Y}{r}= \tilde M \tilde Y   \qquad \text{with} \quad \tilde M=
	\begin{pmatrix}
		 {\cC'}/{\cC} +\cConf '/\cConf & -i\omega^2 + {2 i \lambda \Phi}/{\cC } \\
		- i \Gamma & \Delta +\cConf '/\cConf 
	 \end{pmatrix} \, .
	 \label{firstordertilde}
\end{equation}
 We can now compare this with the system describing the GR dynamics of axial modes in a background 
  \begin{eqnarray}
 \dd{\tilde s}^2 = \tilde{g}_{\mu\nu} dx^\mu dx^\nu = - \tilde \cA(r) \dd{t}^2_* + \frac{1}{\tilde \cB(r)} \dd{r}^2 + \tilde \cC(r)  \dd{\Omega}^2 \,,
 \end{eqnarray} 
 which is characterized, according to \eqref{MatrixMspin2}, by the matrix
 \begin{eqnarray}
\tilde M_{\rm GR} \, =  \, \begin{pmatrix}
		 {\tilde \cC'}/{\tilde \cC} & -i\omega^2 + {2 i \lambda}\tilde \cA/{\tilde \cC } \\
		- i /(\tilde \cA \tilde \cB) & -(\tilde \cA'/\tilde \cA+\tilde \cB'/\tilde \cB)/2
	 \end{pmatrix} \,.
\end{eqnarray}

Remarkably, it is possible to identify the two above matrices $\tilde M$ and $\tilde M_{\rm GR}$, provided one takes
 \begin{eqnarray}
 \label{def_conf}
 \cConf  = \sqrt{ \frac{\Gamma \cB}{\cA}{\cF}^2}\,,
  \end{eqnarray} 
  up to a multiplicative constant (which we have fixed to $1$ without loss of generality). 
 The coefficients of the effective metric are then
 \begin{eqnarray}
 \tilde \cA = \cConf  \, \Phi  \, , \qquad
 \tilde \cB =  \frac{1}{\cConf  \, \Phi \Gamma}\,, \qquad
 \tilde \cC=  \, \cConf  \cC \,,
 \label{tildefuncalpha}
 \end{eqnarray}
so that the full effective metric finally reads
\begin{eqnarray}
 \label{effectivemetric}
 \dd{\tilde s}^2 = \tilde{g}_{\mu\nu} \dd{x^\mu} \dd{x^\nu} = \, \abs{ \mathcal{F}}  \sqrt{ \frac{\Gamma \cB}{\cA}}\left( 
 - \Phi \dd{t_*}\!\!\!^2 + \Gamma \Phi \dd{r}^2 + \cC  \dd{\Omega}^2 \right) \, .
 \end{eqnarray}
 Note that we have imposed a positive $\cConf$ in order to ensure a physically meaningful signature for the effective metric\footnote{Indeed, a negative $\cConf$ immediately yields a  $(--)$ signature in the angular sector.}.
{Moreover, $\Gamma$ is implicitly assumed to be positive in the definition \eqref{def_conf}, which is true only if  $c_*^2$ is positive, according to \eqref{speedcstar1}, meaning there is no gradient instability. 
}

In summary, we have recovered an  effective metric  of the form \eqref{conformalmetric}  but the  conformal factor must  now be fixed (up to an irrelevant constant factor) in order to have the equivalence between the DHOST system and axial perturbations in {\it General Relativity}. Note  that this effective metric agrees with the analysis of  \cite{Tomikawa:2021pca} based on the quadratic action for the perturbations. A related work is   \cite{Takahashi:2021bml}, which  uses a different coordinate system (see App.~\ref{linkwithLemaitre} for the link with our analysis).

\medskip
Finally, let us stress that the equivalence between the DHOST and GR systems can also be seen  in the Schr\"odinger formulation, although it is less transparent. To check this explicitly, one must  compare \eqref{Schrodingerequation}-\eqref{Veff} with the GR Schr\"odinger-like  equation characterized by the choice  $n=\sqrt{\tilde\cA\tilde\cB}$ and the potential \eqref{PotABC} where $\cA$, $\cB$ and $\cC$ are replaced by $\tilde\cA$, $\tilde\cB$ and $\tilde\cC$ respectively.  Imposing the same choice for $n$ in the first equation and identifying the $\lambda$-dependent terms in the two Schr\"odinger-like  equations give two relations\footnote{One gets $\tilde \cB=1/(\Gamma \tilde \cA)$ and $\tilde \cA= \Phi\tilde \cC/\cC$.} that enable us to express $\tilde\cA$ and $\tilde\cB$ in terms of $\tilde\cC$. Substituting this in the GR potential and identifying it with the potential of the DHOST equation, one finally gets
\begin{eqnarray}
\frac{3}{2} \left( \frac{\tilde \cC'}{\tilde \cC}\right)^2 - \frac{\tilde \cC''}{\tilde \cC} + \frac{1}{2} \frac{\Gamma'}{\Gamma} \frac{\tilde \cC'}{\tilde \cC} = 2 \, V_0 \,,
\end{eqnarray}
where $V_0$ is given in \eqref{V_0}. One can verify explicitly that $ \tilde \cC=  \, \cConf  \cC$, with $\cConf$ given in \eqref{def_conf}, is solution of the above differential equation.

 \subsection{Quadratic DHOST: disformal transformations and effective metric}
If we  now restrict our analysis to {\it quadratic} DHOST theories, the coefficients in \eqref{PhiGammaDeltaPhi} reduce to
\begin{align}
\mathcal{F} &= \cA F_2 -  (q^2 + \cA X) A_1 \,,\qquad  
	 \Psi = q \frac{\psi' A_1}{\mathcal{F}}  \,,\nonumber\\
	{\Phi} &=\frac{\mathcal{F}} {F_2 - X A_1}\,, \qquad \Gamma = \Psi^2 + \frac{{q^2 A_1} + \cA {F_2}}{\cA\cB\mathcal{F}} \,.
	\label{PhiGammaDeltaPhi_quad}
\end{align}
In the special case $q=0$, these coefficients further simplify and, substituting them in \eqref{effectivemetric},  the effective metric reduces to 
\begin{eqnarray}
 \label{effectivemetric_quad_q0}
 \dd{\tilde s}^2 = \tilde{g}_{\mu\nu} \dd{x^\mu} \dd{x^\nu} =  \, \sqrt{F_2(F_2 - X A_1)}\left( 
 - \cA\dd{t}\!^2 + \frac{F_2}{F_2 - X A_1} \frac{\dd{r}^2}{\cB} + \cC  \dd{\Omega}^2 \right) \,.
 \end{eqnarray}
One notes that this  is a very simple transformation of the initial background metric, with rescalings that depend  only on $F_2$ and $F_2 - X A_1$. In fact, this transformation can be interpreted  as a disformal transformation, as we now explain, and this remains valid even in the case $q\neq 0$.

By using the correspondence between DHOST theories, via field redefinitions, 
 it is possible to put the coefficient $A_1$ to zero via a disformal transformation of the metric
 \begin{eqnarray}
 \label{disformal}
\hat{g}_{\mu\nu} =\cCo \, g_{\mu\nu} + \cDis \,  \phi_\mu \phi_\nu \, .
\end{eqnarray}
Indeed, any quadratic DHOST action $\hat{S}$ written as a functional of $\hat{g}_{\mu\nu}$ and $\phi$ is related to another quadratic DHOST action $S$ for $g_{\mu\nu}$ and $\phi$, defined by
\begin{eqnarray}
S[g_{\mu\nu}, \phi]\equiv \hat S[\hat g_{\mu\nu}= \cCo \, g_{\mu\nu} + \cDis \,  \phi_\mu \phi_\nu , \phi]  \, .
\end{eqnarray} 
The relations between the quadratic-order coefficients in the respective actions are given by the expressions (see \cite{BenAchour:2016cay}
or Appendix D in \cite{Langlois:2020xbc})
\begin{eqnarray}
\hat F_2 & = & {[{\cCo^2 (1 + X \cDis/\cCo)}]^{-1/2}} F_2 
 \, ,\label{eq:cond-F2} \\
\hat A_1 & = & \left( 1 + X {\cDis}/{\cCo} \right)^{3/2} (A_1 - \frac{\cDis}{\cCo+ \cDis X} F_2) 
\, ,
\end{eqnarray}
and we do not need here the analogous expressions for the other coefficients.
  
It is easy to check that one can impose $\hat F_2=1$ and $\hat A_1=0$
by choosing\footnote{We assume here that $\hat F_2>0$, i.e. $F_2>0$, otherwise the effective Planck mass squared is negative, leading to a theory plagued with ghost instabilities.}
\begin{eqnarray}
\cCo \, =  \, \sqrt{F_2(F_2 - X A_1)} \,, \qquad \cDis \,  = \, \frac{ F_2 A_1}{\sqrt{F_2(F_2 - X A_1)}} \,, 
\label{coeffs_disformal}
\end{eqnarray}
corresponding to the disformed metric
\begin{eqnarray}
	\label{alphaquadratic}
	\hat{g}_{\mu\nu} =   \sqrt{F_2(F_2 - X A_1)} \left(  g_{\mu\nu} +   \frac{A_1}{F_2 - X A_1}  \phi_\mu \phi_\nu \right) \, .
\end{eqnarray}   
In other words, the part of the original DHOST Lagrangian for $g_{\mu\nu}$ that determines the dynamics of  axial perturbations is equivalent to another Lagrangian for $\hat{g}_{\mu\nu}$ where the relevant part is the same as in GR (the other coefficients of the Lagrangian are also modified in the disformal transformation and can remain nonzero, but they are irrelevant for axial perturbations).

By comparing this statement with the result of the previous subsection, it is clear that the effective metric obtained previously should coincide with the disformally related metric for which the dynamics of the axial perturbations is the same as in GR. Let us check this explicitly. The disformal transformation \eqref{disformal} applied to the static spherically metric \eqref{metric}
and scalar field \eqref{generalscalar}
yields
\begin{eqnarray}
\dd{\hat s}^2 = -(\cCo \cA - \cDis q^2)  \left( \dd{t} - \cDis \frac{ q \psi'}{\cCo \cA - \cDis q^2} \dd{r} \right)^2 + \cCo \left( \frac{1}{\cB} + \cDis \frac{ \cA \psi'^2}{\cCo \cA - \cDis q^2}\right)  \dd{r}^2 + 
\cCo\,  \cC  \dd{\Omega}^2 \,.
\end{eqnarray}
Substituting \eqref{coeffs_disformal} and noting that
\begin{eqnarray}
 \sqrt{F_2(F_2 - X A_1)} =  \, \abs{\mathcal{F}} \sqrt{ \frac{\Gamma \cB}{\cA}} \,, 
 \end{eqnarray}
one recovers the effective metric \eqref{effectivemetric}.
Note that the conformal factor in \eqref{alphaquadratic} is well defined if 
\begin{eqnarray}
\label{NogradientQuad}
F_2(F_2-X A_1) >0 \,, 
\end{eqnarray}
which also  guarantees that $\Gamma>0$. 
 This agrees with the no-ghost condition given in \cite{Tomikawa:2021pca} and   \cite{Takahashi:2021bml}.

\subsection{Examples of effective metrics}
 In this subsection,  after some general considerations  on the  causal structure of the background and effective metrics, we compute and discuss the effective metric \eqref{effectivemetric} for the three solutions presented in section \ref{subsection:BH_solutions}.   
 For simplicity, we  restrict our analysis to the  case where the background metric is a black hole with $\cA(r)=\cB(r)$,  which applies to our three examples.
 
 \subsubsection*{Comparison of causal structures}

As mentioned earlier,  non-gravitational fields (e.g. photons, or any type of matter) are minimally coupled to the metric $g_{\mu\nu}$ and  therefore propagate in the background geometry. By contrast,  axial gravitons behave as if they propagate (in the GR sense) in the effective metric $\tilde g_{\mu\nu}$, as we have seen previously.

The fact that gravitational perturbations and other fields effectively ``live" in different  geometries might lead to  interesting new physical effects or inconsistencies.
A simple and straightforward analysis consists in checking that the causal structures associated with the two metrics are compatible, following a similar  analysis\footnote{{In  \cite{Babichev:2018uiw}, the authors  compared the effective metric of the radial scalar perturbation with the physical metric where non-gravitational fields propagate.} }  discussed in  \cite{Babichev:2018uiw}.

According to \eqref{effectivemetric}, the lightcone and the time-like region delimited by it  are defined, for the effective metric, by 
\begin{equation}
    - \Phi (\dd{t} - \cA \Psi \dd{r_*})^2 + \Phi \Gamma \cA^2 \dd{r_*}^2 \leq 0 \,,
    \label{eq:def-lightcone-grav}
\end{equation}
which is equivalent to
\begin{equation}
\Phi\left(\dd{t}-a_+  \dd{r_*} \right)\left(\dd{t}-a_- \dd{r_*}\right)\geq 0\,,
\end{equation}
where we have introduced the coefficients $\aminus$ and $\aplus$ defined by
\begin{equation}
    a_\pm(r) = \cA (\Psi \pm \sqrt{\Gamma}) \,.
\end{equation}
When $\Psi=0$, this reduces $a_\pm(r) =  \pm \cA \sqrt{\Gamma}$ so that the lightcones are symmetric. By contrast, when $\Psi\neq 0$, the lightcones become skewed, as we will see in the stealth BH example.

\subsubsection*{Stealth solutions}
For the stealth solution \eqref{metricstealth} and \eqref{scalarstealth}, the coefficients in \eqref{PhiGammaDeltaPhi} are given by \cite{Langlois:2021aji}
\beq
\begin{aligned}
\label{ABCStealth}
&
\cF=1-\frac{\rg}{r}\,, \quad \Psi = \frac{\zeta \, \rs^{1/2} r^{3/2} }{(r-\rs)(r-\rg)}\,, \quad 
\Phi = \frac{r-\rg}{(1+\zeta)r} \,, \\
&\Gamma = \frac{(1+\zeta)r^2}{(r-\rg)^2}\,, \quad
\Delta = \frac{1}{r} - \frac{1}{r-\rg} \,,
\end{aligned}
\eeq
where we have introduced  the constant parameter
\bea
\label{czetadef}
\rg\equiv (1+\zeta)\rs\,.
\eea

Substituting these coefficients into \eqref{effectivemetric}, one finds that the  effective metric
can be written  as another  Schwarzschild metric:
\begin{eqnarray}
\dd\tilde s^2 \; = \; - \left( 1 - \frac{R_g}{R}\right) \dd{T}^2 +  \left( 1 - \frac{R_g}{R}\right)^{-1} \dd R^2 + R^2 \dd\Omega^2 \, ,
\end{eqnarray}
in the rescaled coordinates $(T,R)$ defined as
\begin{eqnarray} 
R=(1+\zeta)^{1/4} r \,, \quad 
T=(1+\zeta)^{-1/4} t_*\, , \quad 
R_g=(1+\zeta)^{1/4} r_g \, , 
\end{eqnarray}
This metric describes a Schwarzschild solution whose horizon is located at $r=r_g$ and therefore is  shifted with respect to the horizon $\rs$ of the background metric. Note that a similar   double-horizon structure  was previously studied in e.g. \cite{Babichev:2006vx}.
 
\begin{figure}[!htb]
\includegraphics{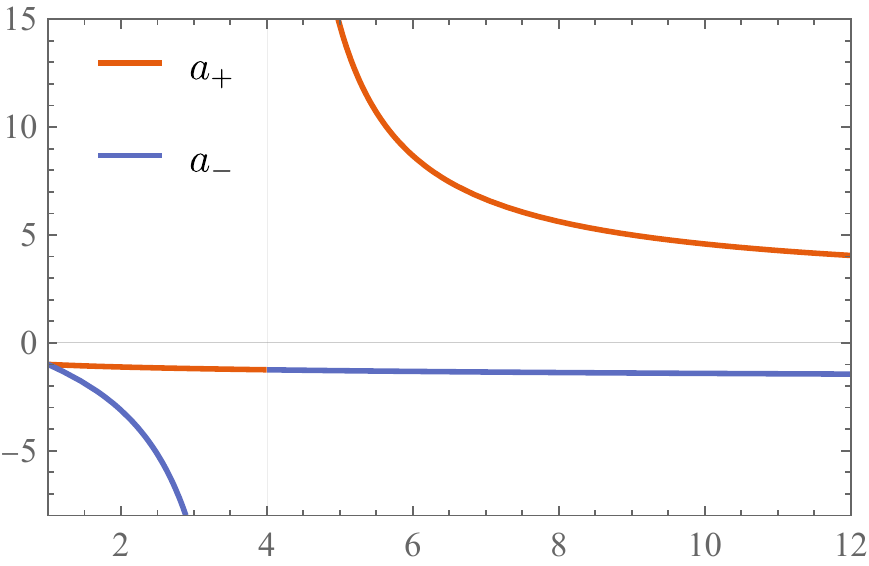}
\caption{Coefficients $a_\pm(r)$ as functions of $r$, with the choice of parameters  $r_s = 1$ and $\zeta = 3$. The vertical line is placed at $r = r_g$.}
\label{fig:a_pm}
\end{figure}

 The coefficients $a_\pm$ associated with the  lightcone of the effective metric are given here by
\begin{equation}
    a_\pm(r) = \cA (\Psi \pm \sqrt{\Gamma}) = \frac{\zeta \sqrt{r\, \rs}}{r - r_g} \pm \sqrt{1+\zeta} \frac{r-\rs}{|r-r_g|} \,.
\end{equation}
and their radial dependence is shown in Fig.~\ref{fig:a_pm}. 

\begin{figure}[!htb]
	\begin{subfigure}[t]{0.45\textwidth}
		\includegraphics{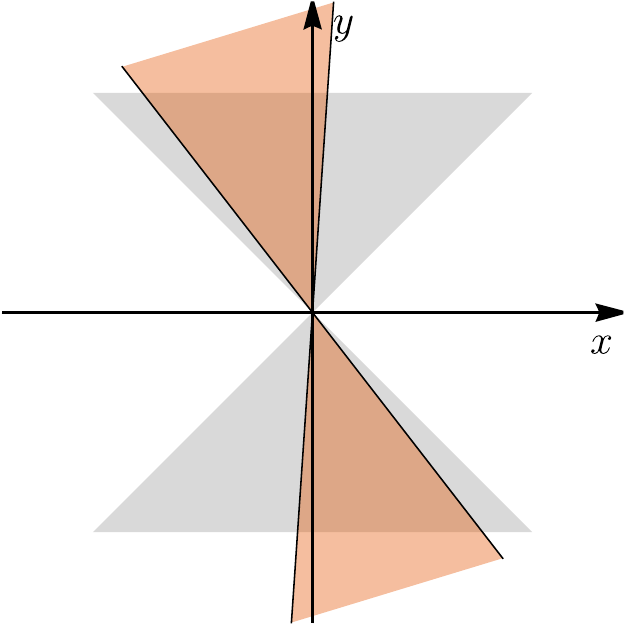}
		\caption{Lightcones for $r > r_g$.}
		\label{fig:lightcones-grav1}
	\end{subfigure}
	\begin{subfigure}[t]{0.45\textwidth}
		\includegraphics{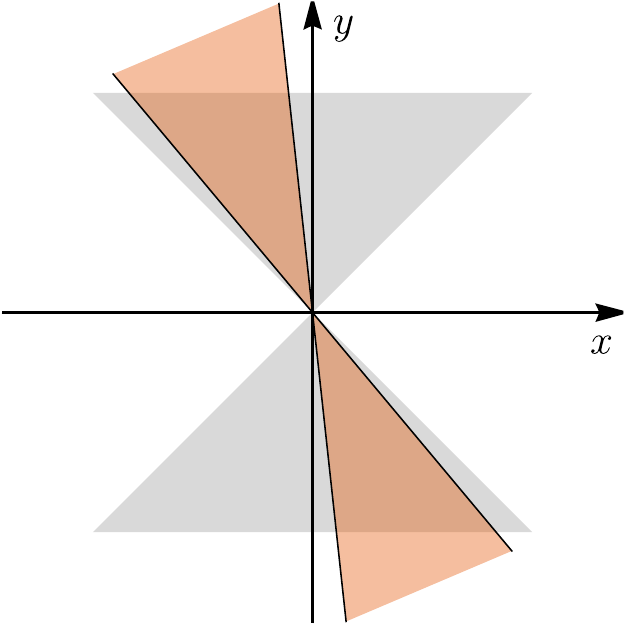}
		\caption{Lightcones for $\rs < r < r_g$.}
		\label{fig:lightcones-grav2}
	\end{subfigure}
	\caption{At some point $(t, r_*)$ in spacetime, we  plot the lightcones in the $(x=\dd{r_*} , y=\dd{t})$ plane  for a stealth black hole with $\rs=1$ and $\zeta=3$. The background lightcones for the background metric are shown in gray. The two cases shown here correspond respectively to $r = 5$ and $r = 3$.}
\label{fig:lightcone}
\end{figure}

We have also plotted the corresponding lightcones, inside and outside the effective horizon $r_g$, in Fig.~\ref{fig:lightcone}. In the same figure, 
the lightcone associated with  the background stealth BH coincides with the standard Minkowski lightcone, since the metric is conformally related to Minkowski in the coordinates $(t, r_*)$ which we are using. We find that the relative position of the lightcones is the same inside and outside $r_g$. This means that  the causal structures are compatible since it is possible to define a common spatial hypersurface on which to specify initial data.

\subsubsection*{BCL solution}
 
We consider now the BCL solution \eqref{BCLrprm} and \eqref{BCLscalar}. The coefficients governing the dynamics of axial perturbations are given by  \cite{Langlois:2021aji}
\bea
&&\cF=\af \cA\,, \qquad  \Psi = 0 \, , \qquad \Phi = \cA \, , \qquad
 \Gamma =\frac{r^2(r^2 + 2 r_+ r_-)}{(r-r_+)^2 (r+r_-)^2} \, ,  \nonumber \\
&&
\Delta = -\frac{r_+}{r(r-r_+)} +  \frac{r_-}{r(r+r_-)} \, . \label{coeffcalBCL}
\eea
This leads to the effective metric 
\begin{eqnarray}
\dd\tilde s^2 \; = \; f_0 \sqrt{1+ \xi \frac{\mass^2}{r^2}} \left[ -\cA(r) \dd t^2+ \frac{1}{\cA(r)} \left( 1+ \xi \frac{\mass^2}{r^2}\right)  \dd r^2+ r^2 \dd\Omega^2  \right]\, ,
\end{eqnarray} 
where the dimensionless parameter $\xi = 2 r_+ r_-/\mass^2$  measures the deviation from General Relativity.  Even though the effective metric  differs from the background metric, it still describes a black hole geometry,  with a  horizon that coincides with the background horizon located at $r=r_+$.
 Hence, the effective metric is regular in the domain $]r_+,\infty[$ and the Schr\"odinger-like equation for axial perturbations  can be solved following the same strategy as in GR.
 
{
Note that the effective lightcones, characterized by
\beq
a_\pm=\pm\sqrt{1+ \xi \frac{\mass^2}{r^2}},
\eeq
are symmetric and always inside the background lightcones, since $|a_\pm|>1$.
In summary,   
}
one does not expect instabilities for 
the axial perturbations contrary to the polar sector in which we found some evidence of instabilities in \cite{Langlois:2021aji}.

\subsubsection*{D$\rightarrow$4 Gauss-Bonnet solution.}

In the D$\rightarrow$4 Gauss-Bonnet solution \eqref{eq:metric-function-r}-\eqref{eq:sol-phi},
the coefficients of the linear system, computed in \cite{Langlois:2022eta}, are given by
\begin{equation}
	\label{F_gamma_i}	\mathcal{F} = \frac{f \gamma_1}{z^2} \,,\quad \Gamma = \frac{\gamma_2}{f^3 \gamma_1 \gamma_3}\,, \quad 
\Phi = \frac{f \gamma_1}{\gamma_3} \,,\quad 
\Delta =  - \frac{\mathcal{F}'}{\mathcal{F}}\,,
\end{equation}
where  we have introduced the functions $f(z)= \sqrt{\cA(r)}$,  with $z=r/r_+$, and 
\begin{eqnarray}
\gamma_1 & = & f \left[ z^2 + 2\beta (f-1)(f-1-2z f')\right] \, , \\
\gamma_2 & = & z^4 - 2\beta (1+\beta)z \, , \\
\gamma_3 & = & z^2+2\beta (1-f^2) \, ,
\end{eqnarray}
with $\beta = \alpha/r_+^2$. This leads to the effective metric
\begin{eqnarray}
\label{effectiveGB}
\dd\tilde s^2 \; = \; - \frac{1}{z^2} \sqrt{\frac{f \gamma_1^3 \gamma_2}{\gamma_3^3}} \, \dd t^2+   
\frac{1}{z^2} \sqrt{\frac{\gamma_1 \gamma_2^3}{f^5 \gamma_3^5}}\,  \dd z^2+ \sqrt{\frac{\gamma_1 \gamma_2}{ f \gamma_3}} \, \dd\Omega^2  \,.
\end{eqnarray} 

 This metric is clearly very different from the background metric and its  behaviour when $z\to 1$ 
 can be determined from the analysis of the functions $\gamma_i$.  
One can see that $\gamma_1$ and $\gamma_3$ are positive when $z>1$ and  $\gamma_2$ is positive when  $z> z_2=[2\beta(1+\beta)]^{1/3}$. In order to have the effective metric \eqref{effectiveGB} well defined  in  the vicinity of $z=1$, we assume that $\beta$ is small enough so that $z_2<1$, i.e. $\beta < (\sqrt{3}-1)/2$. 
In this case, the behaviour of the effective metric in the limit $z\to 1$
takes the form 
\begin{eqnarray}
\dd\tilde s^2 \; \simeq \; -c_1 (z-1)^{1/4}   \dd t^2 + c_2 (z-1)^{-5/4}  \dd z^2+ c_3 (z-1)^{-1/4}  \dd \Omega^2 \, ,
\label{eq:effective-metric-4dEGB-horizon}
\end{eqnarray}
where the $c_i$ are constants, since  the functions $\gamma_i$ tend to a  constant value  while $f(z)\simeq f_0 (z-1)^{1/4}$ with $f_0$ constant.

The Ricci scalar associated with the effective metric \eqref{eq:effective-metric-4dEGB-horizon} behaves as $R \simeq (z-1)^{-3/4}$ in this limit, which indicates that the effective metric describes a naked singularity\footnote{This can be expected in general for BH solutions in DHOST theories with cubic terms, since avoiding a singularity requires a specific tuning of the behaviours of $\Phi$ and $\Gamma$ near the horizon. The complicated structure of these coefficients, given in \eqref{PhiGammaDeltaPhi}, {suggests that such a tuning is unlikely,  unless for very specific DHOST theories.}}. The consequence is that the dynamics of axial modes is very different from the GR case and, in particular,  ingoing and outgoing modes cannot be defined  at the horizon,  as  shown in \cite{Langlois:2022eta}. This
might lead to stability issues associated with spatial divergences of the metric perturbations, which we leave for a  future study.

\section{Conclusion}
In the present work, we have studied the axial perturbations of black hole solutions in the context of DHOST theories, including cubic terms in second derivatives of the scalar field. The main reason to focus our attention on axial perturbations is that they are much simpler than their polar counterparts, thus 
leading to  more general conclusions. 
Indeed, in DHOST theories, the axial perturbations contain a single degree of freedom, as in GR, whereas the polar perturbations, by contrast, contain two degrees of freedom, due to the inclusion of the scalar field perturbation.  Given this relative simplicity, the dynamics of the axial modes, which can be written as a first-order radial system, can be described by a {\it single} Schr\"odinger-like equation. 

Reinterpreting the Schr\"odinger-like equation as a Klein-Gordon equation, one can naturally define, up to an arbitrary conformal factor, an effective geometry in which the perturbations propagate. Remarkably, one can exhibit a direct correspondence between the DHOST axial system and a GR axial system associated with the effective metric, whose conformal factor is now determined. In the case of quadratic DHOST theories, it turns out that this effective metric can be obtained directly from the background metric by a disformal transformation. 

The reason for this surprising property is that the dynamics of axial modes depends on only two independent functions of the DHOST quadratic Lagrangian and these two functions can be mapped into their respective GR values via a disformal transformation that also depends on two free functions. By contrast, for more general DHOST theories which include cubic terms, the dynamics of axial perturbations  depends on four independent functions, which cannot be related to their GR version via a disformal transformation. Nevertheless, one can still define an effective metric, since all information from the five Lagrangian functions and the background metric boil down to just four radially-dependent functions in the dynamical system.  This dynamical system can be mimicked by GR axial perturbations living in an appropriately chosen metric. This metric is artificial and does not necessarily correspond to a GR vacuum solution (but can nevertheless be seen as a solution of Einstein's equations with an artificial energy-momentum whose perturbation does not contribute to the dynamics of axial perturbations).

We have computed the effective metric for three known black hole solutions: a stealth solution, whose metric coincides with the standard Schwarzschild solution, a solution similar to the Reissner-Nordstr\"om solution but with a negative charge squared, and finally a solution obtained in a special 4d limit of Gauss-Bonnet theories. Interestingly, we find that the effective metric is also a BH metric in the first two cases, with a shifted horizon in the stealth case. By contrast, in the last case, the effective geometry, in the region close to the horizon, corresponds to that of a naked singularity.

We have also checked,  for each example, that  the background and effective causal structures are compatible, i.e. that  it is possible to find a common hypersurface where initial data can be specified, by comparing locally the respective lightcones of the two metrics. While the compatibility is immediate for two of our examples, the presence of a shifted  horizon in the stealth case requires more caution and we have verified that, even in the region between the two horizons,  the two metrics remain compatible. These various cases illustrate the diversity of the  effective metrics that one can encounter. It would be interesting to further explore the range of possible effective metrics, depending on the  choices of DHOST functions.

Finally, let us conclude with some comments about future extensions. As mentioned earlier, polar perturbations in DHOST theories are more complicated than axial perturbations as they contain two degrees of freedom. This implies that their dynamics cannot be described by a {\it single} Schr\"odinger-like equation and it is not obvious how the notion of effective metric could be defined in this context. It should however be possible to study the high frequency limit of the perturbations: in this limit, the perturbations would decouple and we expect to be able to define a local effective metric  for each degree of freedom. We plan to explore this direction in a future work. It would also be interesting to study more systematically  DHOST theories with cubic terms.

\begin{acknowledgments}
We would like to thank Eugeny Babichev, Christos Charmousis and Antoine Leh\'ebel for very instructive discussions on effective metrics.
\end{acknowledgments}

\appendix

\section{Axial perturbations and their equations of motion}
\label{Appendix:Axial}
In this appendix, we give more details on how to obtain the equations of motion for  axial perturbations
in a general  cubic theory of the form \eqref{eq:generic-quintinc-Horndeski}, assuming shift symmetry.
At this stage, we do not assume degeneracy. Nonetheless, let us recall that 
 the black hole background solutions we have considered have been computed for DHOST theories.

Interestingly, only the five elementary Lagrangians $L^{(2)}_1$, $L^{(3)}_2$, $L^{(3)}_3$ and $L^{(3)}_6$ contain pieces of the form  $\phi_{\mu \nu} \phi^{\nu \rho}$, which entails that  they are the only ones  contributing to the
dynamics of  axial perturbations. All the other elementary Lagrangians, together with the terms $P(X)$ and $Q(X)\Box \phi$, can be ignored as  far as the dynamics of  axial perturbations is concerned, which  drastically simplifies  the calculations. 

\medskip

In the  Regge-Wheeler gauge in which the  non-vanishing components of
the axial perturbations are
\begin{align}
	&h_{t\theta} = \frac{1}{\sin\theta}  \sum_{\ell, m} h_0^{\ell m}(t,r) \partial_{\varphi} {Y_{\ell m}}(\theta,\varphi), \qquad
	h_{t\varphi} = - \sin\theta  \sum_{\ell, m} h_0^{\ell m}(t,r) \partial_{\theta} {Y_{\ell m}}(\theta,\varphi), \nonumber \\
	&h_{r\theta} =  \frac{1}{\sin\theta}  \sum_{\ell, m} h_1^{\ell m}(t,r)\partial_{\varphi}{Y_{\ell m}}(\theta,\varphi), \qquad
	h_{r\varphi} = - \sin\theta \sum_{\ell, m} h_1^{\ell m}(t,r)  \partial_\theta {Y_{\ell m}}(\theta,\varphi), \label{eq:odd-pert}
\end{align}
which are expanded in spherical harmonics ${Y_{\ell m}}(\theta,\varphi)$.
Since the background metric is static, it is also convenient to work in the frequency domain. In practice, any partial derivative with respect to the time coordinate $t$ corresponds to a multiplication by $-i\omega$.

\medskip

As explained in \cite{Langlois:2021aji}, only two perturbation equations (out of the four non-trivial Einstein equations) are independent.  Hence the system of perturbation equations can be  reduced to a system of two second order equations, which read
\begin{equation}
	\begin{aligned}
		0 &= \omega \fa_1(r) h_0'(r) + \qty(\lambda \fa_2(r) + \omega^2 \fa_3(r)) h_1(r) + \qty(q \lambda \fa_4(r) + \omega \fa_5(r))h_0(r) \,,\\
		0 &= q \fa_6(r) h_0'(r) + \fa_7(r) h_1'(r) + \qty(q \fa_8(r) + \omega \fa_9(r)) h_0(r) + \qty(\fa_{10}(r) + q \omega \fa_{11}(r)) h_1(r) \,,
	\end{aligned}
	\label{systemwithai}
\end{equation}
where $\lambda$ has been defined in \eqref{lambda}. The coefficients  $\fa_i$, whose expressions are too cumbersome to be written here,  are functions of $r$ (but not  of $\lambda$, $\omega$ or $q$) and depend on the Lagrangian of the theory and on the background solution. They  satisfy the following properties:
\begin{eqnarray}
&&\fa_3=i \fa_1 \, , \qquad
\cC \, \fa_5 = - \cC' \, \fa_1 \, , \qquad
\fa_{11} = -i \fa_6 \, , \nonumber \\
&&\fa_2 \fa_6 = \fa_4 \fa_7 \, , \qquad \fa_6 (\fa_{10} - \fa'_7) = \fa_7( \fa_8 - \fa_6') \, ,
\label{relationsai}
\end{eqnarray}
for any choice of functions $F_2$, $A_1$, $F_3$, $B_2$ and $B_6$,  even if they do not satisfy the degeneracy conditions \eqref{eq:degen-DHOST-cubic}, \eqref{degeneracy1} and \eqref{degeneracy2}. 

We now wish to reformulate  the system \eqref{systemwithai}  in  the canonical form
\begin{equation}
	\dv{Y}{r} = (M_0 + \omega M_1 + \omega^2 M_2) Y \,,
	\label{eq:forme-canonique-generale}
\end{equation}
where the components of the vector $Y$ are independent linear  combinations of $h_0$ and $h_1$, and the three matrices 
$M_0$, $M_1$ and $M_2$ do not depend on $\omega$. 
To do so let us try 
the following ansatz:
\begin{equation}
Y =\begin{pmatrix} Y_1 \\ Y_2 \end{pmatrix} \qq{with} Y_1=h_0\,, \quad \omega Y_2 = h_1 - q f h_0  \,,
\end{equation}
where $f$ is an undetermined  function at this stage. \eqref{systemwithai} implies that  the differential system satisfied by $Y$ is given by
\begin{equation}
	\begin{pmatrix}	- \omega \fa_1 & 0 \\ -q(\fa_6 + f \fa_7) & -  \omega \fa_7 \end{pmatrix} \dv{Y}{r} = \begin{pmatrix} q \lambda (\fa_4 + f \fa_2) + \omega (\fa_5 + q \omega f  \fa_3)& \omega (\lambda \fa_2 + \omega^2 \fa_3) \\ q(\fa_8 + f \fa_{10} + f' \fa_7) + \omega (\fa_9 + q^2  f \fa_{11}) & \omega (\fa_{10} + q  \omega \fa_{11})	\end{pmatrix} Y \,.
	\label{eq:systeme-a_i}
\end{equation}
To remove the off-diagonal term in the left-hand side matrix and simplify the system, one chooses
\begin{equation}
f \, = \, - \frac{\fa_6}{\fa_7} \,,
\end{equation}
which also implies that 
\begin{eqnarray}
\fa_4 + f \fa_2=0 \, , \qquad \fa_8 + f \fa_{10} + f' \fa_7 = 0 \, , 
\end{eqnarray}
due to the last two relations in \eqref{relationsai}. Using  the remaining three relations in  \eqref{relationsai}, the  system further reduces to
\begin{eqnarray}
	\dv{Y}{r} = \begin{pmatrix}
		{\cC'}/{\cC} + i \omega \Psi & -i\omega^2 + {2i\lambda}\Phi/{\cC}  \\
		- i \Gamma  & \Delta +  i \omega \Psi 
	\end{pmatrix} Y \,,
	\label{systemwithalpha}
\end{eqnarray}
with
\begin{equation}
	\Psi = q \frac{\fa_2}{\fa_4} \,,\quad \Phi = i  \cC \frac{\fa_2}{2\fa_1} \,,\quad \Gamma = i \frac{ q^2 \fa_2 \fa_{11} - \fa_4 \fa_9 }{\fa_7 \fa_4}  \, , \quad \Delta = -\frac{\fa_{10}}{\fa_7} 
	\, .
	\label{funcGammaetc}
\end{equation}
  Finally, one obtains the formulas given in \eqref{PhiGammaDeltaPhi} by substituting the explicit expressions for the  functions $\fa_i$.

\medskip

Let us note  that the system \eqref{systemwithalpha} describes the dynamics of a single degree of freedom whereas it has been obtained without imposing any degeneracy condition.  Hence, the Ostrogradsky ghost does not show up in the axial sector of the perturbations and should appear when one 
considers polar perturbations. This result is 
fully consistent with the analysis of \cite{Tomikawa:2021pca} based on the computation of the quadratic Lagrangian. Furthermore, it is similar to what happens in the context of cosmological perturbations where the Ostrogradsky ghost can be seen to appear in the scalar sector and not in the tensorial sector  \cite{Langlois:2017mxy}.

 \section{Dispersion relation in the high frequency limit from the first order system}
 \label{dispersionfirstorder}
In this appendix, we 
give another method to compute the effective metric, up to a conformal factor,   directly from the first-order system without relying on the Schr\"odinger-like reformulation 
used in subsection \ref{section:conformal}. 

This method relies on the dispersion relation of the first-order system
\eqref{eq:system-axial-canonical} in the high frequency  and large $\ell$ limits. In this  regime, the components of $Y$ can be viewed as plane waves and one can introduce the wave number $k$ such that
\begin{eqnarray}
\dv{Y}{r}= i k Y \, .
\end{eqnarray} 
Furthermore, $k$, $\omega$  and $\ell$ are supposed to scale in the same way in the large frequency limit. 
Hence, the dispersion relation can be obtained  by  simply requiring   that the  determinant  
\begin{eqnarray}
\text{det} (- i k \mathbb I + M  )  & = &  
\text{det} \left(
\begin{array}{cc}
		 -i k+ \frac{\cC'}{\cC} + i \omega \Psi& -i\omega^2 + \frac{2 i \lambda \Phi}{\cC } \\
		- i \Gamma & -ik + \Delta + i \omega \Psi
	 \end{array} \right)\\
& = &  -(  k -  \omega \Psi)^2 - \frac{\Phi \Gamma}{\cC}\ell^2 + \Gamma \omega^2 + {\cal O}(\omega)  \, ,
\end{eqnarray} 
where $\mathbb I$ is the identity, 
vanishes at  leading order  in $\omega$. 
This corresponds to propagation equation of the form
\begin{eqnarray}
 - \Gamma \pdv[2]{\chi}{t_*}  + \pdv[2]{\chi}{r}  + \frac{\Phi \Gamma}{\cC} \Delta\!^{(2)}  \chi \, \approx 0 \, 
\end{eqnarray}
in this limit, for the quantity $\chi$ defined in \eqref{fullchi}
which is consistent with the Schr\"odinger-like equation \eqref{eq_chi}.

\medskip
One might wonder what happens when working with a different set of functions, i.e. another vector $\hat Y$ in place of $Y$.  Writing the relation between the two vectors as $Y=P\hat Y$, where $P$ is assumed to be independent of  $\omega$, one finds that $\hat Y$  satisfies the new  first-order system 
\begin{eqnarray}
\dv{\hat Y}{r}= \hat M \hat Y \, , \qquad \hat M = P^{-1} M P - P^{-1} \dv{P}{r} \, .
\end{eqnarray}
Such change of vector  is ubiquitous in \cite{Langlois:2021aji} where we study the asymptotic behaviour of perturbations.

When $P$ is an arbitrary invertible matrix, the $\omega$-dependency of $\hat M$ is generically very different from the $\omega$-dependency of $M$. More precisely, $\hat M$ will still have the same $\omega$-structure as $M$, i.e.
\begin{eqnarray}
\hat M = \hat M_0 + \omega \hat M_1 + \omega^2 \hat M_2 \, ,
\end{eqnarray} 
where $\hat M_1=M_1=i \Psi \mathbb I$   and $\hat M_2$ is nilpotent but not necessarily in its Jordan form.  
 Hence, in general, one does not expect  the dispersion relation obtained from $\hat M$,
 \begin{eqnarray}
 \text{det} (-i k \mathbb I + \hat M) \, = \, 0 \, ,
 \end{eqnarray} 
 to be  equivalent to the one computed previously from $M$. 
One can check that, for this new dispersion relation to be valid,
$\hat M_2$ must be  its Jordan block form after the change of variables.

\section{Stealth solution in  Lema\^itre coordinates}
\label{linkwithLemaitre}

In this appendix, we  discuss how  some of our results, in particular the propagation speed,  are  related to those of the recent work \cite{Takahashi:2021bml}, in which  the stealth Schwarzschild  solution
\begin{eqnarray}
	\cA(r)=\cB(r)= 1 - \frac{\rs}{r} \, , \qquad \phi(r,t)= q t + \psi(r) \, , \quad \psi'(r)= q \frac{\sqrt{r \rs}}{r - \rs} \, ,
\end{eqnarray} 
where $\rs$ is the Schwarzschild  radius and $X=-q^2$, 
is described in   Lema\^itre coordinates $(\tau,\rho)$, defined by
\begin{eqnarray}
	\label{gcoordinates}
	\mathrm \tau = \ \mathrm  d t + \alpha(r) \mathrm d r \, , \qquad
	\mathrm \rho = \mathrm  d t + \beta(r) \mathrm d r \, ,
\end{eqnarray}
with 
\begin{eqnarray}
	\alpha = \frac{\sqrt{r \rs}}{r - \rs} \, , \qquad
	\beta = \frac{r^2}{\sqrt{r \rs}(r-\rs)} \,.
	\label{eq:chgmt-var-lemaitre}
\end{eqnarray}
The authors of  \cite{Takahashi:2021bml} computed the quadratic Lagrangian for the axial perturbation $\chi$ and obtained an associated equation of motion, which  is not   Schr\"odinger-like. Indeed, one can check, starting from our Schr\"odinger-like equation,  that only a uniform linear combination of our coordinates $t_*$ and $r_*$ conserves the Schr\"odinger form that we have derived, 
similarly to the Lorentz transformations of special relativity.

The equation satisfied by the perturbation $\chi$ is given by~\cite{Takahashi:2021bml}
\begin{eqnarray}
	- \frac{\partial }{\partial \tau}\left( s_1  \frac{\partial \chi}{\partial \tau} \right) + \frac{\partial }{\partial \rho}\left( s_2  \frac{\partial \chi}{\partial \rho} \right)  + W \chi \, = \, 0 \, ,
\end{eqnarray}
where 
\begin{eqnarray}
	s_1= \frac{(1+\zeta)^2 r^6}{\sqrt{\rs/r}} \, , \qquad s_2 = \frac{(1+\zeta) r^6}{(\rs/r)^{3/2}} \, ,
\end{eqnarray}
and the expression of $W$, which  is not needed here,   can be found in \cite{Takahashi:2021bml}. Notice that $r$ depends on 
$\tau$ and $\rho$, hence the equation is time-dependent. 

Even if this equation does not have the canonical wave equation form, one can  deduce the associated propagation speed from its high frequency limit, which amounts to take into account only the derivatives of  highest order, i.e. second order here.  
This gives 
\begin{eqnarray}
\label{cL}
	c_{\rm L}^2 = \frac{s_2}{s_1} = \frac{r}{\rs (1+\zeta)} \,.
\end{eqnarray}
In fact, the propagation speed  given in \cite{Takahashi:2021bml}   is defined by
the relation
\begin{eqnarray}
	c_\rho^2 = -\frac{g_{\rho\rho}}{g_{\tau \tau}} c_{\rm L}^2 \, ,
\end{eqnarray}
in order to express the propagation speed in normalised units (i.e. in the normalised basis spanned by the vectors $e_\tau=(-g_{\tau \tau})^{1/2} \partial_\tau$, $e_\rho=(g_{\rho\rho})^{-1/2}\partial_\rho$).

Now, we would like to  relate the propagation speed \eqref{cL} to our relation \eqref{speedcstar1}
\begin{eqnarray}
	c_*^2 = \frac{1}{n^2 \Gamma} = \frac{(r-r_g)^2}{(1+\zeta) (r-\rs)^2} \qq{with} \zeta = 2q^2 F_X \, , \qquad r_g = (1+\zeta) \rs \,,
\end{eqnarray}
\medskip
also obtained  in \cite{Langlois:2021aji},  using the usual tortoise coordinate with $n(r) = \cA(r)$. 
Starting from the wave operator in \eqref{Schrodingerequation}, one finds, after restoring  the time derivatives and applying the coordinate change \eqref{gcoordinates},
\begin{eqnarray}
	\frac{1}{n^2} \left(\pdv[2]{}{r_*} - \frac{1}{c_*^2} \pdv[2]{}{t_*}  \right)& = &\left( \tilde \beta^2- \frac{1}{n^2 c_*^2}\right)\pdv[2]{}{\rho} + \left(  \tilde \alpha^2 - \frac{1}{n^2 c_*^2} \right)  \pdv[2]{}{\tau} + 2 \left( \tilde \alpha \tilde \beta - \frac{1}{n^2 c_*^2} \right)  \pdv[2]{}{\tau}{\rho}  \nonumber  \\
	& +&  \left( \tilde \alpha' + \frac{n'}{n}\tilde \alpha \right)  \pdv{}{\tau} +   \left( \tilde \beta' + \frac{n'}{n}\tilde \beta\right)  \pdv{}{\rho} \, ,
	\label{newschroop}
\end{eqnarray}
where we have introduced the coefficients
\begin{eqnarray}
	\tilde \beta = \beta +  \Psi \, , \qquad
	\tilde \alpha = \alpha + \Psi \, .
\end{eqnarray}
One can then check that the cross derivative term in \eqref{newschroop} vanishes and the propagation speed squared deduced from the first two terms is 
\beq
c^2_{\rm new}= \frac{\tilde\beta}{\tilde\alpha}\,,
\eeq
which coincides with the expression \eqref{cL}.

We have thus checked that the two different expressions for the propagation speed given in  \cite{Langlois:2021aji} and \cite{Takahashi:2021bml} agree, up to an adequate change of coordinates. One can note that, independently of their expressions, the absence of gradient instability requires the same condition  $1+\zeta > 0$.

\bibliographystyle{utphys}
\bibliography{full_biblio}

\end{document}